\documentclass[pra,aps,twocolumn,nofootinbib,floatfix,showpacs]{revtex4}
\usepackage{epsfig}
\usepackage{graphicx}
\usepackage[usenames,dvipsnames]{color} 
\usepackage{psfrag}
\usepackage{graphics}
\usepackage{hyperref}
\usepackage{color}
\begin{document}

\title{Collective oscillations in  ultra-cold atomic gas}

\author{J.T. Mendon\c{c}a$^{1,2}$}
\email{titomend@ist.utl.pt}
\author{R. Kaiser$^3$}
\author{H. Ter\c{c}as$^2$}
\author{J. Loureiro$^1$}

\affiliation{$^1$CFP and $^2$CFIF, Instituto Superior T\'{e}cnico, Av. Rovisco Pais 1, 1049-001 Lisboa, Portugal}
\affiliation{$^3$Institut Non Lin\'eaire de Nice, UMR 6618,  1361 Route des Lucioles, F-06560 Valbonne, France}

\begin{abstract}

Using both fluid and kinetic descriptions, where repulsive forces between near by atoms are included, we discuss the basic oscillations and waves of a cloud of ultra-cold atoms confined in a magneto-optical trap. The existence of a hybrid mode, with properties similar to both plasma and acoustic waves is described in detail. Tonks-Dattner resonances for confined hybrid modes in a spherical cloud are discussed and the prediction of a nonlinear coupling between the dipole resonanc and the hybrid modes is considered. Landau damping processes and quasi-linear diffusion in velocity space are also discussed.
\end{abstract}

\maketitle

\section{Introduction}

In recent years, and mainly due to the emergence of laser cooling techniques \cite{cooling} there has been an increasing interest on the physics of ultra-cold atoms. This interest has been mainly concentrated on the study of Bose Einstein condensates, and on the theoretical and experimental understanding of their properties  \cite{legget, dalfovo}. However, attention has also been given recently to the study of collective oscillations of non-condensed atomic molasses in the magneto-optical traps \cite{kaiser,kaiser2,distefano}. From this work, collective behavior similar to that observed in plasma physics, has emerged, leading to the discovery of an equivalent electric charge for the neutral atoms, and of electrostatic type of interactions between nearby atoms \cite{walker}, and to Coulomb like explosions of the atomic cloud when the magnetic confinement is switched off \cite{pruvost}. The theoretical modeling of such collective processes is still not well established, and this is the main motivation of the present work. Here we propose to apply the well known methods of waves and oscillations in plasmas to such processes in a neutral gas. 

In this work we consider the collective behavior of ultra-cold atom gas, in order to  identify the basic mechanisms of oscillation, to establish its frequencies and to derive dispersion relations for its basic modes of propagation phenomena. We will use both fluid and kinetic descriptions where the main forces associated with the laser cooling processes are retained. 

This paper is organized as follows: in Sec. II, we define some essential parameters to describe the features of a cold atomic cloud confined in a magneto-optical trap. The basic set of fluid equations is also established. In Sec. III, we report the existence of brand new hybrid mode for short wavelength scales, somewhere between plasma waves and acoustic waves. In Sec. IV, we adress the oscillations for the long wavelength scales, where we discover modified Tonks-Dattner resonances, which correspond to confined hybrid oscillations inside the atomic cloud, formally similar to those in ref. \cite{parker}. Nonlinear hybrid resonances driven by the dipolar oscillations are predicted in Sec. V. We also use a wave kinetic approach, in order to refine the fluid description and describe new phenomena. Hence, in Sec. VI we describe the system via a Fokker-Planck equation that can be directly derived from the Schroedinger equation for the collective field of the atoms in the radiation and trapping fields. It is shown that the waves and oscillations in the gas can be damped by a resonant atomic interaction with the collective oscillations, which is a manifestation of Landau damping. We will also show that a spectrum of collective oscillations will lead to diffusion in the atomic velocity space, thus preventing the laser cooling process to proceed further. This effect is added to other diffusion processes already known in the literature. Finally, some conclusions are stated.

\section{Basic equations and fluid description} 

The simplest model to describe a gas of cold atoms in a magneto-optical trap is based on Doppler cooling \cite{Haensch1975, Wineland1975} 
and on the spatial confinement due to the presence of a magnetic field gradient. The relevant average forces acting on a single atom 
are based on the quasi-resonant radiation pressure force and 
can be written 
(for a convenient choice of the relative
polarisations of the laser beams and the atomic transition)
for each direction $r_i$ ($r_1=x$, $r_2=y$, $r_3=z$) as 

\begin{equation}
\begin{array}{cc}
\displaystyle{F_i\left( r_i,v_i \right) = \frac{\hbar k_L \Gamma }{2} s_{inc} \left[ \frac{\Gamma^2}{\Gamma^2+ 4(\Delta- \mu_i r_i - k_L v_i)^2}\right.}\nonumber\\
\displaystyle{\left.- \frac{\Gamma^2}{\Gamma^2+ 4(\Delta+ \mu_i r_i +k_L v_i)^2} \right]}
\end{array}
\label{force} 
\end{equation}
This expression relies on the low intensity Doppler model for the magneto-optical force 
(incident on-resonance 
saturation parameter per beam $s_{inc}=I_0/I_{sat}\ll 1$). The Zeeman shifts (described by $\mu_i r_i$) and Doppler shifts ($k_L v_i$) are 
responsible for trapping and cooling respectively. 
Here $I_0$ is 
the laser intensity of laser beams incident along the six directions, $\Gamma$ the natural line width  
of the transition used in the cooling process and $\Delta$ the 
frequency detuning between the laser frequency $\omega_L = k_L c$ and the atomic transition frequency $\omega_{at}$.
Assuming symmetric 
forces ($\mu_i=\mu$) along each of the three directions $Ox$, $Oy$ and $Oz$ one can write to first order in $\vec{r}$ and $\vec{v}$

\begin{equation}
\vec{F}_{MOT} = - \kappa \vec{r} - \alpha \vec{v},
\label{eq:2.1a} 
\end{equation}
where $\kappa$ is the spring constant of the trap and $\alpha$ the friction coefficient, 
 
\begin{equation}
\alpha= - 8 \hbar k_L^2 s_{inc}\frac{\Delta/\Gamma}{\left[1+4\Delta^2/\Gamma^2\right]^2}\quad,\quad \kappa = \alpha \mu / k_L 
\label{eq:2.1b} 
\end{equation}
which is related to the dipole frequency

\begin{equation}
\omega_d=\sqrt{\frac{\kappa}{M}}\label{trapping}
\end{equation}
where $M$ represents the mass of a single atom. 
Even though most magneto-optical traps using for instance alkaline atoms (such as Na, Li, K, Rb or Cs) are not completely described by 
such a simple Doppler model \cite{cooling}, 
the forces above constitute a good first approach to describe the dynamics of single atoms in a magneto-optical trap.

This description of a magneto-optical trap is known to be limited to only moderate number of atoms (typically $10^5$). For larger atom number 
additional forces need to be taken into consideration. 
A second force to be considered is the shadow force, or absorption force, $\vec{F}_A$, and was  first discussed by Dalibard \cite{dalibard}. 
This is associated with the gradient of the incident laser intensity due to laser absorption by 
the atomic cloud. It is an attractive force which can be determined by a Poisson type of equation, as

\begin{equation}
\nabla \cdot \vec{F}_A = - \frac{\sigma_L^2 I_0}{c} n (\vec{r})
\label{eq:2.2a} \end{equation}
where $n (\vec{r})$ is the atom density and $\sigma_L$ the laser absorption cross section. 
Finally, a third force, $\vec{F}_R$, can be called repulsive force or radiation trapping force, and was first considered 
by Sesko et al \cite{sesko}. It  describes atomic repulsion, due to the radiation pressure of scattered photons 
on nearby atoms, and can also be determined by a Poisson type of equation

\begin{equation}
\nabla \cdot \vec{F}_R = \frac{\sigma_R \sigma_L I_0}{c} n (\vec{r})
\label{eq:2.2b} \end{equation}
where $\sigma_R$ is the atom scattering cross section. A detailed discussion of these forces and explicit expressions for the 
cross sections $\sigma_R$ and $\sigma_L$ can be for instance found in \cite{sesko, pruvost}.
These expressions for the forces acting on the atomic clouds and due to the laser cooling beams, correspond to the simplest possible description of 
the laser cloud interaction, and can be used in a first approximation to model the fluid dynamics of the ultra-cold gas, which can be derived by computing the zeroth and the first momenta of the Fokker-Planck equation, neglecting the diffusion term \cite{pruvost}. The basic set of equations can then be written as

\begin{equation}
\frac{\partial n}{\partial t} + \nabla \cdot ( n \vec{v} ) = 0
\label{eq:2.3a} \end{equation}
\begin{equation}
\frac{\partial \vec{v}}{\partial t} + \vec{v} \cdot  \nabla \vec{v} = - \frac{\nabla P}{M n} + \frac{\vec{F_T}}{M}
\label{eq:2.3b} \end{equation}
where $n$ and $\vec{v}$ are the density and velocity field of the gas respectively, $\vec{F_T}=\vec{F}_{MOT}+ \vec{F_c}$ and $P$ is the gas pressure. Here we have also defined the collective force $\vec{F_c} = \vec{F}_A + \vec{F}_R$, which is determined by the Poisson equation resulting from equations (\ref{eq:2.2a}) and (\ref{eq:2.2b}), 

\begin{equation}
\nabla \cdot \vec{F_c} = Q n \quad , \quad Q = (\sigma_R - \sigma_L) \sigma_L I_0 / c
\label{eq:2.4a} \end{equation}
The system can then be regarded as a one component plasma where the electrostatic attractive force due to ions can be formaly replaced by the confining force, in such way that, for the unperturbed state $n=n_0$, we may represent the total force through a Laplace-like equation

\begin{equation}
\nabla\cdot \vec{F_{T}}(n=n_0)=0.
\label{eq:2.4b}
\end{equation}
In typical experimental conditions the repulsive forces largely dominate over the shadow effect, and the quantity $Q$ is positive \cite{walker,pruvost,kaiser}, which allows us to define the typical frequency
\begin{equation}
\omega_P=\sqrt{\frac{Qn_0}{M}},
\label{freqp}
\end{equation}
which is a straight-forward generalization of the well-known electron plasma frequency. Comparing this expression with the usual definition of the plasma frequency $\omega_{pe}$ in an ionized medium, we conclude that neutral atoms behave as if they had an equivalent electric charge, as first  noticed by \cite{walker}, with the value $q_{eff} = \sqrt{\epsilon_0 Q}$, where $\epsilon_0$ is the vacuum electric permitivity. The experimental value observed for this effective atomic charge is $q_{eff} \sim 10^{-4}$ to $10^{-6}$ times the electron charge. 
%%%%%%%%%%%%%%%%
In a typical MOT experiment, we expect to have $n_0\approx 10^{10}\mbox{cm}^{-3}$, $M\approx 10^{-25} \mbox{Kg}$, what yields a plasma frequency in the range of $\omega_P/2\pi=q_{eff}/2\pi\sqrt{m_e/M}\omega_{pe}\approx100~ \mbox{Hz}$.
%%%%%%%%%%%%%%%%%%
It is clear from equation (\ref{freqp}) that plasma like oscillations are only possible for $Q > 0$. Therefore, they cannot occur when the shadow force (\ref{eq:2.2a}) dominates over the repulsive scattering force (\ref{eq:2.2b}).

\par In order to conclude the analogy between the fundamental parameters between the plasma and the cold atoms, one remark should be made concernig the motion of the center-of-mass
\begin{equation}
\vec{R}=\sum_{i=1}^N\vec{r_i}m_i/M,\label{CM2}
\end{equation}
where $M=\sum_{i=1}^Nm_i$. Therefore, for a typical neutral plasma with spherical geometry, it is a well know result that the center-of-mass oscillates at the so-called Mie frequency $\omega_M=\omega_{pe}/\sqrt{3}$, where the factor $\sqrt{3}$ arises from the spherical symmetry \cite{guerra} and is an essential parameter in the description of resonances in clusters \cite{mulser}. However, for the case of a spherical cloud of cold atoms, it is a simple task to verify that the center-of-mass obeys the following equation of motion

\begin{equation}
\frac{d^2\vec{R}}{dt^2}+\omega_d^2\vec{R}=0,\label{CM}
\end{equation}
which states that the $\vec{R}$ oscillates exactly at the dipole frequency $\omega_d$, corresponding to the generalized plasma frequency $\omega_P$ in the unperturbed state $n=n_0$, as given by eq. (\ref{freqp}). The reason for this difference remains in the fact that, in the case of the spherical plasmas, the restoring force depends on the charge balance between ions and electrons $n_i-n_e$, which on the other hand depends on the shape of the cloud. In the case of cold atoms, the potential is unequivocally determined by the laser and the magnetic field (say trapping) parameters, and therefore the shape and the atomic cloud plays no role. \par 
In what follows, we make use the set of fluid equations derived in this section for a quasi-continum medium and describe the main collective modes. We start from the infinite medium approximation, to emphasize the nature of the oscillations, and then introduce in Sec. IV the effect of the finite size of the cloud. 
   
\section{Plasma hybrid waves}

We first assume oscillations that can be excited in the cold gas with a wavelength much smaller that its radius. The medium can therefore be assumed as infinite.  We then assume that the equilibrium state of the gas is perturbed by oscillations with frequency $\omega$ and wavevector $\vec{k}$. In the sense of linear response theory, we linearize the above fluid and Poisson equations, by defining perturbations around the equilibrium quantities

\begin{equation}
n=n_0+\tilde n,\quad \vec{F}=\vec{F_0}+\delta \vec{F},\quad \vec{v}=\vec{v_0}+\delta\vec{v}.\label{eq:2.5b}
\end{equation} 
Since the trapping force $\vec{F}_{MOT}$ defines only the equilibrium quantities and plays no role in the modes we are about to describe, we drop the subscript $c$ for the perturbation in the collective force $\delta\vec{F_c}$ in (\ref{eq:2.5b}) for the sake of simplicity. For the closure of the system of fluid equations, one equation of state for the hydrodinamical pressure must be given. In this paper, we assume that $P$ is given by an adiabatic equation of the form 

\begin{equation}
P(n) \sim n^\gamma,\label{pressure}
\end{equation}

\begin{figure}
      \includegraphics[angle=0,scale=0.75]{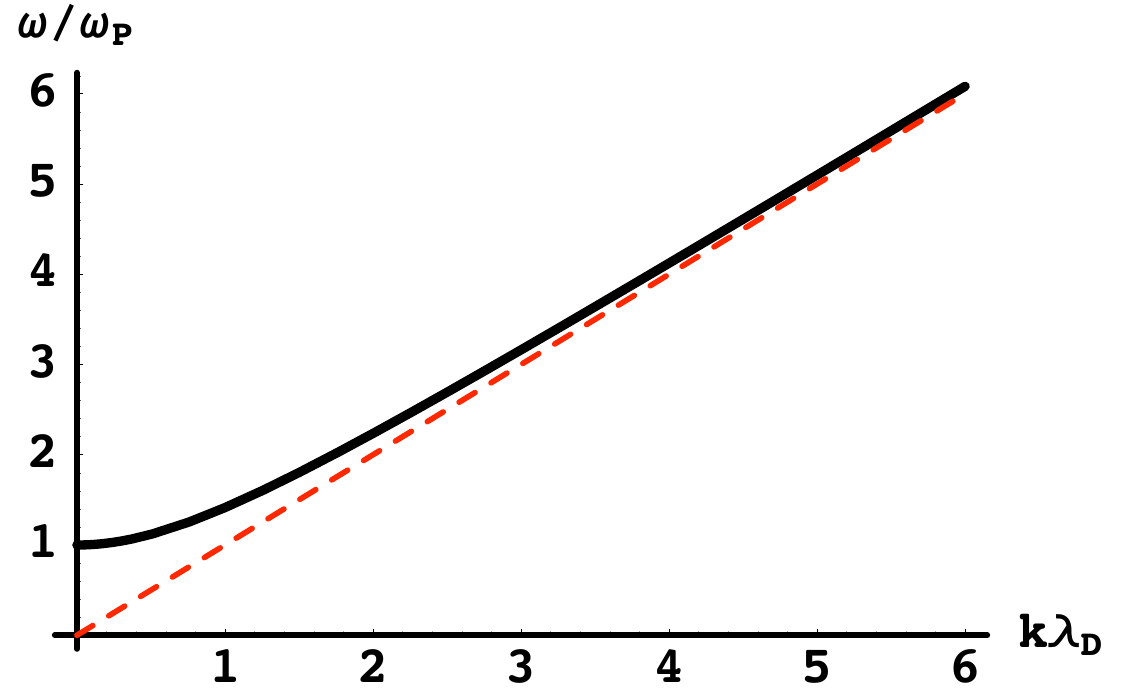}
      	\caption{\label{fig1} (color online) Normalized dispersion relation $\omega/\omega_P$ plotted against $k\lambda_D$. The red dashed line represents the acoustic assimptotic behavior of the hybrid waves.}
\end{figure}

where $\gamma$ represents the adiabatic constant. The implications of this assumption will be stated in Sec. V, in the context of a kinetic description. Using the later together with equations (\ref{eq:2.3a}), (\ref{eq:2.3b}) and (\ref{eq:2.4a}), we can easily obtain
\begin{equation}
\left[ \frac{\partial}{\partial t} \left( \alpha +  \frac{\partial}{\partial t} \right) + \omega_P^2 - u_S^2 \nabla^2 \right] \tilde{n} = \left( \frac{u_S^2 \nabla \tilde{n}}{n_0} - \frac{\delta \vec{F}}{M} \right) \cdot \nabla n_0
\label{eq:2.6} \end{equation}
where $u_S$ can be identified with the sound speed

\begin{equation}
u_S^2 = \gamma \frac{P_0}{M n_0}
\label{eq:2.7} \end{equation}
and $P_0$ is the equilibrium gas pressure.

\par We will now assume that the atomic equilibrium density $n_0$ is uniform, in consistence with the harmonic nature of the trapping potential, therefore neglecting the right hand side of equation (\ref{eq:2.6}). The influence of boundary conditions and inhomogeneities on the collective oscillations of the gas will be discussed later. Assuming a space-time dependence of the perturbations $\tilde{n}$ and $\delta \vec{F}$ of the form $\exp (i \vec{k} \cdot \vec{r} - i \omega t)$, with a complex frequency $\omega = \omega_r + i \omega_i$, we obtain for the dispersion relation and for the corresponding damping rate, the values

\begin{equation}
\omega_r^2 = \omega_P^2 + k^2 u_S^2 + \frac{3}{4} \alpha^2 \quad , \quad  \omega_i = \frac{\alpha}{2}
\label{eq:2.8} \end{equation}
In the limit of very small viscosity $\alpha \ll \omega_P$, the later dispersion relation reduces to $\omega^2 = \omega_P^2 + k^2 u_S^2$, which is formally identical to the dispersion relation of electron or plasma waves in ionized media (also known as Langmuir waves) but where the electron thermal velocity $v_{the} = \sqrt{k_BT_e / m_e}$, ($T_e$ and $m_e$ are the electron temperature and mass) is replaced by the sound velocity divided by a numerical factor $u_S / \sqrt{3}$. 
%%%%%%%%%%%%%%%%%%%%%
According to the experimental parameters refereed in the previous section, we estimate the sound velocity to value $u_S\propto\sqrt{k_BT/M}\approx 20 ~\mbox{cm/s}$.
%%%%%%%%%%%%%%%%%%%%%%%%%%
This shows that the wave mode described by equation (\ref{eq:2.8}) contains elements of both electron plasma waves and acoustic waves. It possesses a lower cut-off, given by $\omega_r =\sqrt{ \omega_P^2 + 3 \alpha^2 / 4}$, which is typical on an electron plasma wave, but its phase velocity tends to the sound velocity $u_S$ and becomes weakly dispersive as an acoustic wave. Its corresponding quasi-particles can therefore be seen as hybrid entities, somewhere between plasmons and phonons. 
%%%%%
The excitation of such modes in a typical MOT setup could be driven by modulating one of the six trapping laser beams. Such modulation should be nearly resonant with $\omega_P$, and hence one may excite hybrid waves by sweeping the modulation frequency around $100 ~\mbox{Hz}$. Associated with the propagation of such waves, we expect to observe a periodic variation on the intensity of the luminosity. By measuring the period of this oscillation, one can identify the excited mode.
%%%%

The existence of such an hybrid mode is one of the main results of this paper. 

\section{Modified Tonks-Dattner resonances}

The hybrid mode discussed above is only meaningful in infinite and homogeneous media. In physical terms, its dispersion relation can only be applied to waves that propagate locally, with wavelength scales much smaller than the inhomogeneity scale and the cloud dimensions. 
Let us consider now oscillations with a wavelength that is comparable with the size of the atomic cloud. In this case we can no longer neglect the boundary conditions. Going back to equation (\ref{eq:2.6}), we assume that the density perturbations oscillate at a frequency $\omega$ as previously, but the corresponding spatial structure will be determined by the expressions

\begin{equation}
\begin{array}{cc}
\displaystyle{\left[ \nabla^2 + k^2 (\vec{r}) \right] \tilde{n} = \frac{\delta \vec{F}}{M u_S^2} \cdot \nabla n_0 + \nabla \ln n_0 \cdot \nabla \tilde{n}},\nonumber\\\\
 \displaystyle{\nabla \cdot \delta \vec{F} = Q \tilde{n}}
\end{array}
\label{eq:3.1} 
\end{equation}
where the space dependent wavenumber $k (\vec{r})$ is defined by 

\begin{equation}
k^2 (\vec{r}) = [\omega^2 - \omega_P^2 (\vec{r})] / u_S^2
\label{eq:3.2} \end{equation}
Before going into a more complex model, it is useful to consider the simple one-dimensional problem \cite{boyd}. In the case of a uniform slab of cold gas, we have $\nabla n_0 = 0$, except at the boundaries $x = 0$ and $x = L$. Equations (\ref{eq:3.1}) and (\ref{eq:3.2}) then reduce to a simple one-dimensional equation 

\begin{equation}
\frac{d^2 \tilde{n}}{d x^2} + \frac{1}{u_S^2} [\omega^2 - \omega_P^2 (x) ] \tilde{n} = 0
\label{eq:3.3} \end{equation}
Taking the boundary conditions $\tilde{n} (0) = \tilde{n} (L) = 0$, we obtain the following dispersion relation

\begin{equation}
\omega_\nu^2 = \omega_P^2 \left[1 + \left( \pi\nu\frac{\lambda_D}{L}\right)^2 \right]
\label{eq:3.4} 
\end{equation}
where the quantum number $\nu$ can take the values $0, 1, 2, 3, ...$, and the quantity $\lambda_D = u_S / \omega_P$ is the Debye length for a cold neutral gas, in analogy with the plasma definition (where however the sound speed $u_S$ is replaced by $\sqrt{3} v_{the}$, as mentioned before). This defines a natural length above which plasma effects should be expected. 
%%%%%%%%%%%%%%%
Following the previously estimated values for $\omega_P$ and $u_S$, we expect to observe a Debye length of the order of $\lambda_D\approx 100~\mu\mbox{m}$. In a typical MOT experiment, the radius of the cloud varies between $a\approx 1-5 ~\mbox{mm}$, yielding the relation $\lambda_D/a\ll 1$. Hence, plasma-like effects are expected to take place in a cloud containing a moderate number of atoms. 
%%%%%%%%%
As a remark, we should stress out that there is no experimental evidence so far of any equation of state $P(n)$, which may compromise the definition of the sound speed $u_S$.\\

\par The relation (\ref{eq:3.4}) shows that the finite dimensions of the slab imply the existence of a series of resonant modes with an integer number of half-wavelengths. The cylindrical geometry was considered, for the plasma case, in a famous paper by Parker, Nickel and Gould in 1964 \cite{parker}, but it is more natural here to consider a spherical geometry for the ultra-cold gas which, to our knowledge, was not derived for a plasma. We expect to find an infinite series of resonances, similar to equation (\ref{eq:3.4}), known as the Tonks-Dattner resonances \cite{tonks,dattner}.
For this purpose, we consider the internal oscillations in a spherical cloud with radius $a$ in the homogeneous case, where $\nabla n_0 (r) = 0$, for $0 \geq r < a$, for which analytical solutions can be found. 
%%%%%%%
These results remain qualitatively valid even for a more realistic density profile, as shown in Ref. \cite{parker} , for the case of a inhomogeneous cylindrical plasma. The case of a non-uniform spherical plasma will be addressed in a future work.
%%%%%

Performing a separation of variables, we can obtain solutions of the form

\begin{equation}
\tilde{n} (\vec{r}) = R (r) Y (\theta, \phi)
\label{eq:3.5} \end{equation}
where $(r, \theta, \phi)$ are spherical coordinates. After separation of variables, we get the usual spherical harmonics for the angular part of the density perturbation

\begin{equation} 
Y (\theta, \phi) = P_l^m (\cos \theta) \exp (i m \phi),
\label{eq:3.6} \end{equation}
where $P_l^m (\cos \theta)$ are the associated Legendre polynomials, $l$ is a positive integer or zero, and $| m | < l$. The radial equation resulting from (\ref{eq:3.1}) and (\ref{eq:3.5}) can be written as

\begin{equation}
\frac{d}{d r} \left( r^2 \frac{d R}{d r} \right) + [ k^2 r^2 - l (l + 1) ] R = 0.
\label{eq:3.7} \end{equation}
By using a simple transformation of variables, $x = k r$, and $S (k r) = \sqrt{k r} \; R (r)$ this equation can be reduced to a Bessel equation

\begin{equation}
\frac{d^2 S}{d x^2} + \frac{1}{x} \frac{d S}{d x} + \left[ 1 - \frac{(l + 1/2)^2}{x^2} \right] S = 0
\label{eq:3.8} \end{equation}
The solutions with regular behavior at the origin $x = 0$ are therefore given by Bessel functions of the first kind, $J_{l + 1/2} (x)$.
From this we conclude that the Tonks-Dattner modes in a spherical homogeneous cold atom cloud are determined by 
  
\begin{equation}
\tilde{n} (\vec{r}, t) = \sum_{l,m}{\tilde{n}_l (t)} \frac{J_{l+1/2} (k r)}{{\sqrt{k r}}} P_l^m (\cos \theta )\exp (i m \phi)
\label{eq:3.9} \end{equation}
where $\tilde{n}_l (t)$ have small amplitudes such that $| \tilde{n}_l | \ll n_0$. The mode frequencies can be obtained by remarking that $\tilde n$ should vanish at the border $r=a$. This implies that the allowed values for $k$ have to obey the condition $k = z_{\nu, l} / a$, where $z_{\nu, l}$ represents the $\nu$th zero of the Bessel function of order $(l + 1/2)$. We are then led to the mode frequencies

\begin{figure}
\includegraphics[angle=90,scale=0.7]{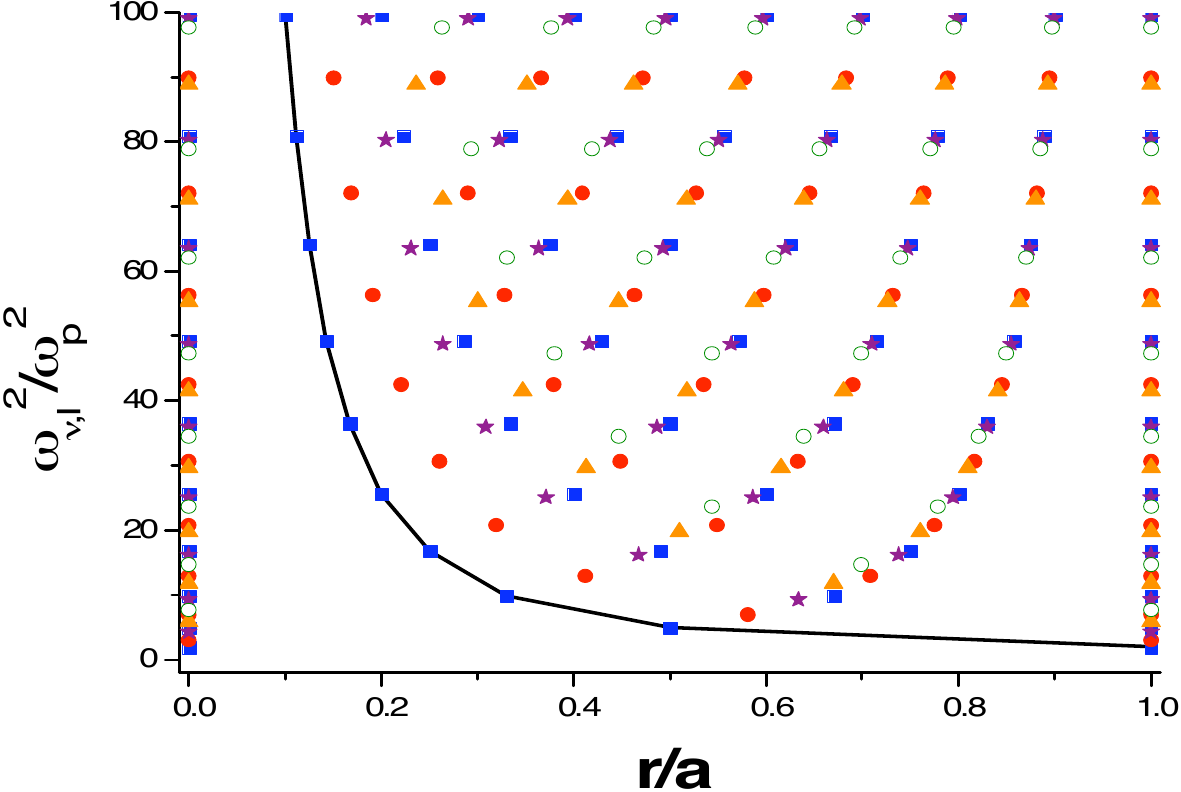}
	\caption{\label{fig2} (color online) Normalized modes $\omega^2_{\nu,l}/\omega_P^2$ plotted against the nodes $r_{\nu,l}/a$ of the radial solution for the density perturbation $\tilde {n}$, for $\lambda_D/a=0.1$ and $1<\nu<10$. Blue squares ($l=0$), red circles ($l=1$), violet stars ($l=2$), yellow triangles ($l=3$), green open circles ($l=4$). The full line is plotted at $l=0$ and scales as $1/\nu^2$.}
\end{figure}

\begin{equation}
\omega_{\nu,l}^2 = \omega_P^2\left\{ 1+ \left(z_{\nu, l} \frac{\lambda_D}{a}\right)^2\right\}.
\label{eq:3.10} \end{equation}
Comparing with the rectangular case of equation (\ref{eq:3.4}) we see that the allowed eigen-frequencies for a spherical cloud now depend on two quantum numbers $\nu$ and $l$. But, in contrast with the similar quantum mechanical solutions for hydrogen like atoms, we have no hierarchical relation between these quantum numbers. The normalized radial profiles for the lowest order solutions are illustrated in figure (\ref{fig2}), 
%%%%%%
which is formally similar to the spectrum that was presented in the past by Dattner for the case of a plasma cylinder \cite{dattner}. Also, a recent experimental on ultracold plasma was published showing the evidence of Tonks-Dattner modes excited by a radio-frequency electric field \cite{fletcher}.

 In a more realistic description, the present rigid (Dirichlet) boundaries will eventually have to be replaced by soft boundaries and a 
generic density profiles $n_0 (r)$ must be assumed, for which numerical solutions have to be found \cite{hugo}. 

\section{Nonlinear oscillations driven by dipolar resonances}

Due to the intrinsic nonlinearity of the fluid description the collective behavior of the medium, it is possible to couple dipole oscillations of the cloud with the hybrid modes. Going back to the fluid equations, and assuming an oscillating mean velocity of the form $\vec{v}_0 \sin (\omega_d t + \phi)$ for the center of mass, we obtain

\begin{equation}
\begin{array}{cc}
\displaystyle{\frac{\partial^2 \tilde{n}}{\partial t^2} + n_0 \nabla \cdot \frac{\partial \vec{v}}{\partial t}}\\
\displaystyle{+ \vec{v}_0 \cdot \nabla \left[ \cos (\omega_d t + \phi) \nabla \tilde{n} + \sin (\omega_d t + \phi) \frac{\partial \tilde{n}}{\partial t} \right] = 0}.
\end{array}
\label{eq:4.5} 
\end{equation}
Using the factorization $\tilde{n}  (\vec{r}, t) = \tilde{A}(t) \tilde{n} (\vec{r})$, the coupling between the center of mass and the hybrid modes can be approximately described by the canonical Mathieu equation 

\begin{equation}
\frac{\partial^2 \tilde{A}}{\partial \tau^2} + [\Delta + 2 \epsilon \cos (2 \tau) ] \tilde{A} = 0,
\label{eq:4.6} \end{equation}
where we have used

\begin{equation}
\tau = \frac{\omega_d}{2} t, \quad
\Delta = \frac{4}{\omega_d^2} (\omega^2 - k^2 u_S^2), \quad
\epsilon = \frac{2}{\omega_d} \vec{v}_0 \cdot \nabla \ln \tilde{n}
\label{eq:4.7a} \end{equation}
It is well known that such an equation has unstable regions. For $| \epsilon | \ll 1$, the first transition to the instability zone occurs for $\Delta \simeq 1 + \epsilon$. We therefore expect to observe an instability of the hybrid and Tonks-Dattner modes, driven by dipole oscillations for

\begin{equation}
\omega^2 \simeq \frac{\omega_d}{2} \left[  \frac{\omega_d}{2} +  \vec{v}_0 \cdot \nabla \ln \tilde{n} \right]
\label{eq:4.7b} \end{equation}
This simple discussion demonstrates the existence of very interesting nonlinear collective phenomena in a cold atom gas. But in the present work we are mainly focused on the linear properties of the medium. A more rigorous and detailed study of nonlinear coupling between dipole and Tonks-Dattner oscillations will be left to future work. 

\section{Kinetic dispersion relations}
\begin{figure}
\includegraphics[angle=0,scale=0.7]{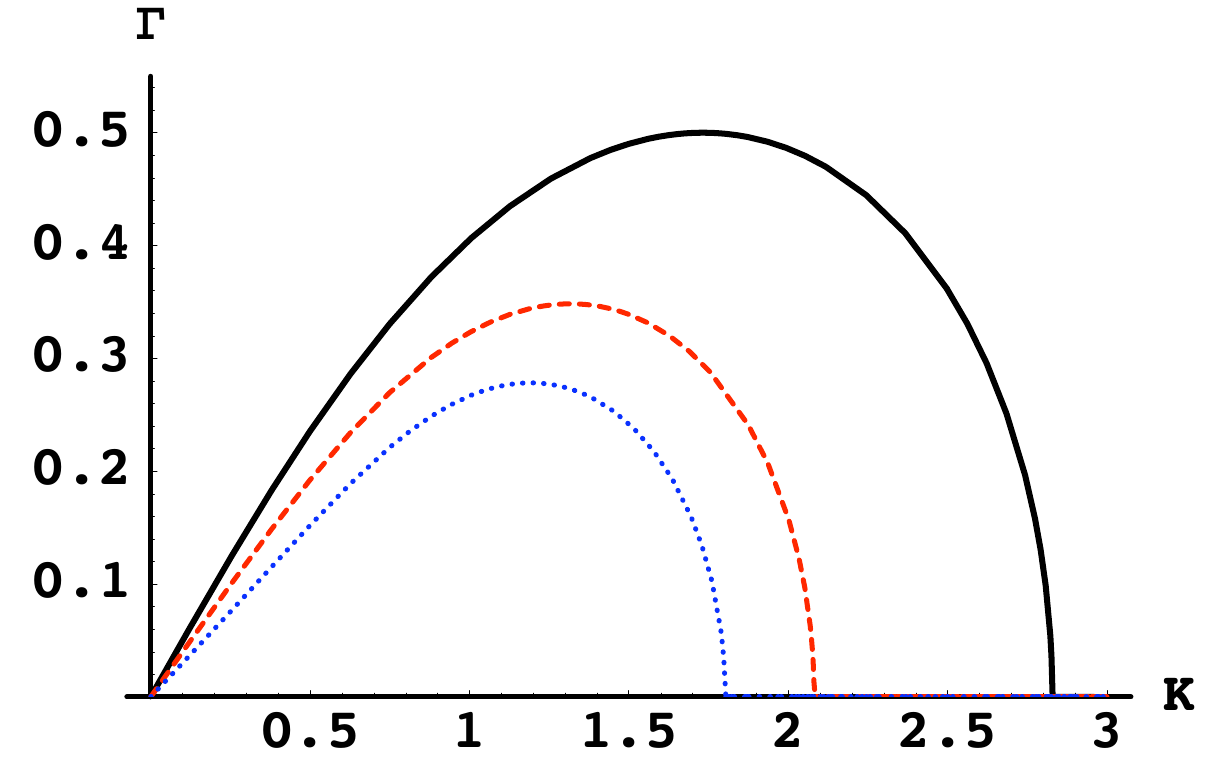}
\includegraphics[angle=0,scale=0.7]{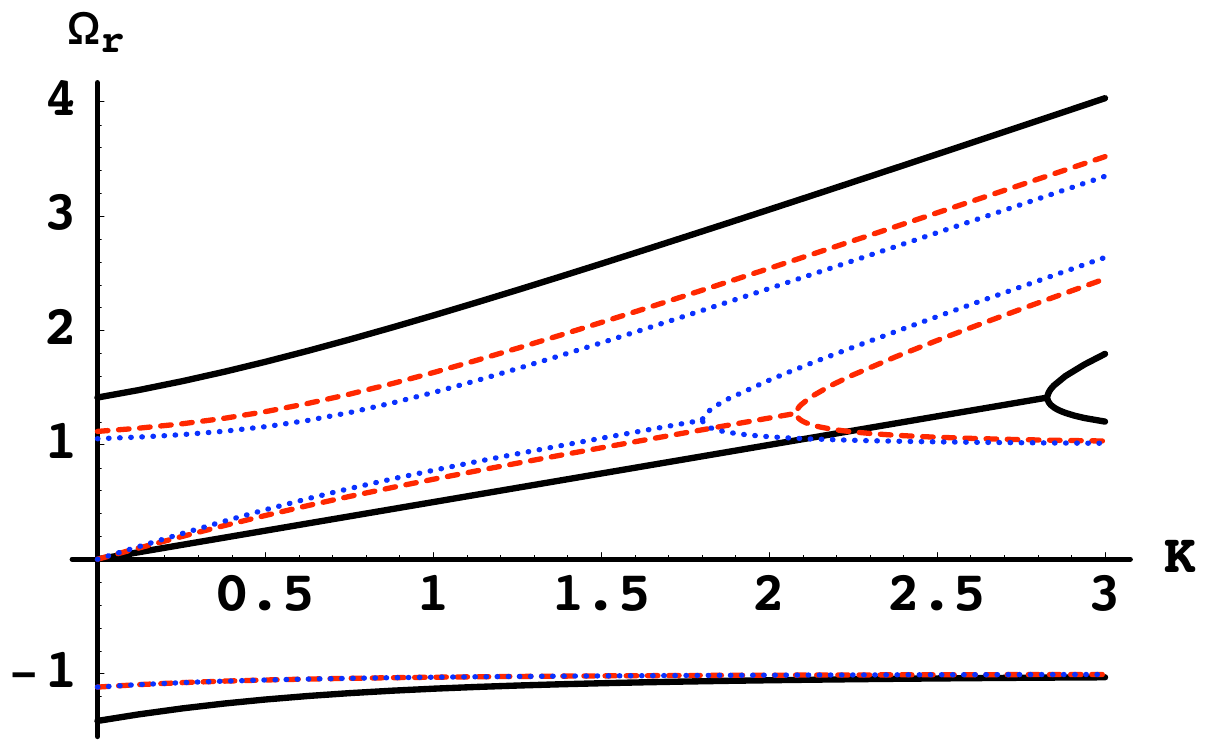}
	\caption{\label{fig3} Normalized roots of Eq. \ref{eq:5.7b} in terms of the dimensionless quantities $\Omega=\omega/\omega_p$, $K=\vec{k}\cdot\vec{v}_1/\omega_p$ and $N=n_1/n_0$. \textbf{(a)} Dimensionless growth rate $\Gamma(K)=+\Omega_i(K)$ (unstable solutions) and \textbf{(b)} dimensionless modes $\Omega_r(K)$ (stable solutions). In both plots, $N=1$ (full black line), $N=1/2$ (dashed violet line) and $N=1/3$ (dotted blue line).}
\end{figure}

%%%%%%%%%%%%%
The analysis of collective oscillations in a cold gas can be refined by using a kinetic description based on the classical probability distribution function $W(\vec{r},\vec{q},t)$. 
%%%%%%%%
This description will allow us to include resonant kinetic processes, which enhance the energy exchanges between part of the atomic population and the hybrid modes excited in the medium. Moreover, a kinetic approach avoids the use of an equation of state for the hydrodinamical pressure $P$. It is rather determined by the statistics, as we will see in the following discussion.
%%%%%%%%%%%%
In the presence of diffusion, the probability distribution function $W(\vec{r},\vec{q},t)$ obeys the Fokker-Planck equation \cite{dalibard2} 

\begin{equation}
\left( \frac{\partial}{\partial t} + \frac{\hbar \vec{q}}{M} \cdot \nabla \right) W = - \frac{\partial}{\partial\vec{k}} [ \vec{F}_{tot} W] + \sum_{ij} D_{ij} \frac{\partial^2 W}{\partial q_i \partial q_j}
\label{eq:5.2} \end{equation}
%%%%%%%%%%%%%%% 
where the total force $\vec{F}_{tot}$ includes the radiative and the damping forces, and the diffusive tensor $D_{ij}$ is due to the fluctuations of the radiative force and spontaneous emission, as discussed by several authors \cite{dalibard2,stenholm}. Here, for simplicity, and because we want to focus on the oscillating modes, we neglect the diffusion term. Diffusion effects will be not completely ignored, rather they will reappear later in a different context. We are then led to a kinetic equation of the Vlasov type, with a damping correction, as given by

\begin{equation}
\left( \frac{\partial}{\partial t} + \vec{v} \cdot \nabla + \frac{\vec{F}}{M} \cdot \frac{\partial}{\partial \vec{v}} \right) W = - \alpha (\vec{v}) W
\label{eq:5.3} \end{equation}
where $\vec{v} = \hbar \vec{q} / M$ is the atom velocity, and the collective (shadow minus repulsive) force $\vec{F}_c$ is determined by a new Poisson equation,  which can be written as

\begin{equation} 
\nabla \cdot \vec{F}_c = Q \int W (\vec{r}, \vec{v}, t) d \vec{v}  
\label{eq:5.4} \end{equation}
This equation is obvious identical to (\ref{eq:2.4a}), because the integral is nothing but the density $n (\vec{r}, t)$, the normalization of the Wigner function. On the other hand, by taking the average of the first two momenta of the kinetic equation (\ref{eq:5.3}), we will be able to derive fluid equations such as (\ref{eq:2.4a})-(\ref{eq:2.4b}). Notice that, the parameter $\alpha$ appearing in the fluid equations in an averaged value of the quantity $\alpha (\vec{v})$ appearing in the wave kinetic equation (\ref{eq:5.3}). In order to focus our attention on the purely kinetic processes, we will assume that $\alpha = 0$ but, at the end of this section, we will discuss the the influence of a finite value for this parameter. We now consider some equilibrium state $W_0 (\vec{v})$ and assume a sinusoidal perturbation, such that $\delta \vec{F}$ and $\tilde{W}$ evolve in space and time as 
$\exp (i \vec{k} \cdot \vec{r} - i \omega t)$. In what follows, we drop the subscript $c$ in the collective force, in consistence with the previous calculations. After linearization, the two previous equations reduce to

\begin{equation}
\tilde{W} = - \frac{i}{M} \frac{\delta \vec{F} \cdot \partial W_0 / \partial \vec{v}}{(\omega - \vec{k} \cdot \vec{v})}
\quad , \quad i \vec{k} \cdot \delta \vec{F} = Q \int \tilde{W} (\vec{v}) \; d \vec{v}
\label{eq:5.5} \end{equation}
From here we get  the dispersion relation for collective cold atom oscillations with frequency $\omega$ and wavevector $\vec{k}$

\begin{equation}
1 + \frac{Q}{M k^2} \int  \frac{ \vec{k} \cdot \partial W_0 / \partial \vec{v}}{(\omega - \vec{k} \cdot \vec{v})} = 0
\label{eq:5.6} \end{equation}
This is similar to that of electrostatic waves in unmagnetized plasmas, and can be rewritten as $1 + \chi (\omega, \vec{k}) = 0$, where the quantity $\chi (\omega, \vec{k}) $ is the susceptibility. In order to understand the physical implications of such a dispersion relation, let us consider first a simple mono-kinetic atomic distribution, of the form $W_0 (\vec{v}) = n_0 \delta (\vec{v} - \vec{v}_0)$, corresponding to a beam of atoms with density $n_0$ and velocity $\vec{v}_0$. In this case, equation (\ref{eq:5.5}) reduces to

\begin{equation}
1 - \frac{Q n_0}{M (\omega - \vec{k} \cdot \vec{v}_0)^2} = 0
\label{eq:5.7a} \end{equation}
This is nothing but the Doppler shifted plasma oscillations discussed above. For $\vec{v}_0 = 0$, this reduces to $\omega = \omega_P \equiv (Q n_0 / M)^{1/2}$. Second, we consider the case of two atomic populations, one at rest with density $n_0$, and another moving with velocity $\vec{v}_1$ with density $n_1 \ll n_0$. This is described by $W_0 (\vec{v}) = n_0 \delta (\vec{v}) + n_1 \delta (\vec{v} - \vec{v}_1)$. The corresponding dispersion relation is given by
 
\begin{equation}
1 - \frac{Q n_0}{M \omega^2} - \frac{Q n_1}{M (\omega - \vec{k} \cdot \vec{v}_1)^2} = 0
\label{eq:5.7b} \end{equation}
The interest of this new dispersion relation is that it leads to an instability, which means that the cold atoms can oscillating spontaneously at a frequency close to $\omega_P$. The energy source driving such oscillations is provided by the atomic beam with density $n_1$. This can be seen by assuming a complex frequency $\omega= \omega_r + i \omega_i$. The maximum growth rate predicted by equation (\ref{eq:5.7b}) corresponds to the resonant case where $\omega_r = \vec{k} \cdot \vec{v}_1$, and is determined by

\begin{equation}
\omega_i = \frac{\sqrt{3}}{2} \; \omega_P \; \left( \frac{n_1}{2 n_0} \right)^{1/3}
\label{eq:5.8} \end{equation}
This is very similar to an electrostatic beam-plasma instability, but can also be seen as the dynamical analogue of the collective atom recoil laser (carl) \cite{bonifacio}, where the incident laser beam is replaced by the atomic beam, and where the resulting stimulated emission is not made of photons, but  of hybrid quasi-particles. The critical region, where unstable modes may show up, is given by

\begin{equation}
K<(1+N^{2/3})^{3/2},
\end{equation}
where we define the dimensionless quantities $K=\vec{k}\cdot\vec{v}_1/\omega_p$, $\Omega=\omega/\omega_p$ and $N=n_1/n_0$. In Fig.\ref{fig3}, we plot the growth rate $\Gamma(K)=\omega_i/\omega_p$ for the unstable region given above and the real part of the spectrum $\Omega_r=\omega_r/\omega_p$, both given by the roots of the dispersion relation \ref{eq:5.7b}. We observe that the stable modes bifurcate exactly at the vanishing point of the unstable ones. \par

Finally, we consider a generic equilibrium quasi-distribution $W_0 (\vec{v})$. The atomic susceptibility can be split into its real and imaginary parts, $\chi = \chi_r + i \chi_i$. By using the Kramers-Kroenig relation between $\chi_r$ and $\chi_i$, and assuming that most of the atoms have velocity smaller than the phase velocity $\omega / k$, which is a very plausible assumption for a gas of ultra-cold atoms, we obtain

\begin{eqnarray}
\chi_r (\omega, \vec{k}) = - \frac{1}{\omega^2} (\omega_P^2 + k^2 u_S^2),\nonumber\\
\chi_i (\omega, \vec{k}) = i \pi \frac{Q}{M k^2} \left( \frac{\partial G_0}{\partial v} \right)_{\omega/k}
\label{eq:5.9} 
\end{eqnarray}
where $v$ is the parallel component of the atom velocity, and we have identified the  sound speed with the integral

\begin{equation}
u_{S}^{2} = \frac{1}{n_0} \int W_0 (\vec{v}) v^2 d \vec{v} = \frac{1}{n_0} \int G_0 (v) v^2 d v
\label{eq:5.10} \end{equation}
The quantity $G_0 (v)$ introduced here is the average of the Wigner distribution over the perpendicular velocities. The later result avoids the postulation of the equation of state (\ref{pressure}). However, since the hydrodynamical pressure 

\begin{equation}
P(\vec{r},\vec{v},t)=\frac{1}{3}M n(\vec{r},\vec{v},t)\left(\langle v^2\rangle-\langle v\rangle^2 \right)
\end{equation}
is a statistical variable we can make use of eqs. (\ref{eq:5.10}) and (\ref{eq:2.7}) to write
\begin{equation}
\gamma=3.
\end{equation}
This means that the hybrid modes are essentially a one-dimensional process, if we remind that $\gamma=(2+d)/d$, which is a well known result for the electron plasma waves. This justifies a posteriori the hybrid character of this oscillations. Back to eq. (\ref{eq:5.7b}), we can easily obtain the dispersion relation, by using $1 + \chi_r (\omega_r, \vec{k}) = 0$, which coincides with equation (\ref{eq:2.8}), and  the wave damping defined by the expression

\begin{equation}
\omega_i = - \frac{\chi_i (\omega_r, \vec{k})}{(\partial \chi_r / \partial \omega )_{\omega = \omega_r}} = \frac{\pi}{\omega} \frac{Q}{M k^2} \left( \frac{\partial G_0}{\partial v} \right)_{\omega/k} \label{eq:5.11} \end{equation}
This is a non dissipative wave damping, which is not related with any increase of the entropy of the physical system, which is the so-called Landau damping. It describes the resonant interactions between the wave and the atomic population with has a parallel velocity nearly equal to the wave phase velocity. Usually, for a thermal equilibrium distribution $W_0 (\vec{v})$ this quantity is negative, and corresponds to wave damping. But the sign of $\omega_i$ can change for a non-thermal distribution, eventually leading to wave instability and wave growth.

We can take a step further in the kinetic description of the collective oscillations in the cold atom cloud, and consider a broad spectrum of fluctuations, described by the total wave intensity

\begin{equation}
I (t) = \int  I(\vec{k}, t) \frac{d \vec{k}}{(2 \pi)^3} 
\label{eq:5.12} \end{equation}
where the spectral intensity is defined by $I (\vec{k}, t) = \tilde{W}^* (\vec{k}, t) \tilde{W} (\vec{k}, t)$. Following the usual steps of the plasma quasi-linear theory, and adapting it to the present context, we can say that, each spectral component behaves in accordance with the above description, and evolves in time according to the equation

\begin{equation}
\frac{d}{d t}  I (\vec{k}, t) = 2 \omega_i (\vec{k}, t) I (\vec{k}, t)  + S (\vec{k}, t) 
\label{eq:5.13} \end{equation}
where $S (\vec{k}, t)$ is any given source term, and the total damping rate $\gamma_k (t)$ slowly evolves in time due to the slow time evolution of the equilibrium (or quasi-equilibrium) distribution $W_0 (\vec{v}, t)$, which can only be considered constant in a short time scale. The temporal evolution of $W_0 (\vec{v}, t)$ under the influence of the fluctuation spectrum is determined by a diffusion equation of the form 

\begin{equation}
\left( \frac{\partial}{\partial t} + \vec{v} \cdot \nabla + \frac{\partial}{\partial \vec{v}} \cdot \mathbf{D} \cdot \frac{\partial}{\partial \vec{v}} \right) W_0 (\vec{v}, t) = 0
\label{eq:5.14} \end{equation}
where the new diffusion tensor $\mathbf{D}$, associated with the collective oscillations, is determined by

\begin{equation}
\mathbf{D}(\vec{v}, t) = \frac{\omega_P^2}{n_0^2} \int I (\vec{k}, t) \frac{\vec{k} \otimes\vec{k}}{(\omega - \vec{k} \cdot \vec{k})} \frac{d \vec{k}}{(2 \pi)^3}
\label{eq:5.15} \end{equation}
Comparing this with our previous kinetic equation (\ref{eq:5.2}) it can be seen that the existence of a collective spectrum of oscillations introduces an additional diffusion effect in atomic velocity space, which tends to prevent the atomic cooling process. As we have noticed, these results are only valid in the limit of a negligible viscosity parameter, $\alpha \rightarrow 0$. A finite value of $\alpha$ will have two distinct consequences. First, it will lead to the damping coefficient already stated by equation (\ref{eq:2.8}), adding to the purely kinetic Landau damping. Second, it will broaden of the Landau resonance appearing in the dispersion relation (\ref{eq:5.6}), and in the diffusion coefficient (\ref{eq:5.15}) therefore reducing the efficiency of the resonant atom-collective wave interactions associated with the Landau resonance and to the quasi-linear diffusion. It will therefore compete with the kinetic effects described in this section. The combined influence of viscous and kinetic damping is outside the scope of the present work and would deserve a separate investigation.

\section{Conclusions}

In this work we have used both fluid and kinetic equations to describe the collective oscillations in a cloud of neutral atoms confined in a magneto-optical trap. Our approach is based on a simple but physically relevant model for the forces acting on the cold atoms, which has been well verified by the experiments, and can be described by a Poisson equation similar to that describing electrostatic interactions \cite{sesko,pruvost,kaiser}. Once the physical picture was established, we started by setting the basic equations and parameters. We have also shown that the existence of internal forces leads to the existence of collective waves which have an hybrid character, with properties that are common to both electron plasma waves and acoustic waves. The similarities and differences of the cold atom gas oscillations and those of a plasma were discussed. 

Taking into account the finite size effect in a cloud of cold atoms, we have shown that internal resonances of these hybrid oscillations inside the cloud can be excited. These modes should be called Tonks-Dattner resonances, in analogy with similar plasma physics effects, since they are shown here also to exist in a neutral magneto-optical trapped gas as well. Previous models developed for plasmas were limited to planar and cylindrical geometries, and they were extended to the spherical geometry which is more relevant to the cold atom clouds. We have extended our discussion of these oscillations to the nonlinear regime, where coupling between dipole oscillations and hybrid waves can take place. Our approximate description suggests that such a nonlinear coupling can lead to the destabilization of hybrid waves or Tonks-Dattner resonances, driven by dipolar oscillations of the cloud center of mass. 

Our analysis proceeded with a kinetic approach. Using this more refined description of oscillations, we were able to derive more general dispersion relations where non-dissipative Landau damping was included. We have limited our discussion to hybrid wave modes with wavelengths much smaller than the typical dimensions of the atomic cloud. But, the results can easily be extended to the confined Tonks-Dattner resonances. Finally, we have established a quasi-linear kinetic equation, showing the occurrence of diffusion in atomic velocity space. This diffusion effect is a direct consequence of the collective fluctuation spectrum, and implies the existence of new collective processes preventing the occurrence of  laser cooling. 

In the present work we have explored the similarities of the could atom cloud with a plasma, which can be associated with the existence of an effective electric charge for the neutral atoms. The resulting wave modes however are not identical to the plasma wave modes, but show an hybrid character. We hope that this work will motive future experimental and theoretical work on the collective oscillations in cold atom traps, and will contribute to launch cold atom research in new directions.

\end{document}